# Study of the Electronic and Structural Properties of Small Molybdenum Clusters via Projector Augmented Wave Pseudopotential Calculations

**Byeong June Min**

*Department of Physics, Daegu University, Kyungsan 712-714, Korea*

We studied the structural and the electronic properties of small Mo clusters (n = 2 ~ 8) via projector augmented wave pseudopotential calculations using plane wave basis functions. Our results show that the 4s- and 4p-semicore states play important roles in the description of small Mo clusters. Also, the dimerization tendency observed in previous theoretical calculations is significantly reduced when the semicore states are explicitly included as the valence states.

## I. INTRODUCTION

Small clusters exhibit novel properties that are different from those of their solid state counterpart. Furthermore, their structural, electronic, and magnetic properties vary sensitively depending on their size, offering a wide possibility of application. For these reasons, small clusters are intensely researched.

Mo is an important constituent in high-strength alloys [1-4]. Mo is also one of the most intensely researched catalyst. More than 50 Mo-containing enzymes are known to exist in bacteria [3-4]. Mo appears commonly in bacterial catalysts used in the breaking of the chemical bond in the nitrogen molecule, achieving biological nitrogen fixation.

Aside from these important applications, Mo is also an interesting research subject in its own. The half-filled electronic d-shell maximizes the atomic spin and promotes strongly directional bonding. For this reason, Mo dimer has once been used as the testing ground for the local spin density approximation (LSDA) [5]. It was a pleasant surprise that these early theoretical investigations turned out to be quite successful [6-7]. However, there exists an unusually large variation in the predictions from density functional theory (DFT). For example, Mo dimer bond length varies very much from 1.65 Å to 2.1 Å with no apparent tendencies that may be attributed to the methodology. Another intriguing question is whether there exists a strong tendency to form dimers in small Mo clusters as reported by Zhang *et al.* [8] and Aguilera-Granja *et al*. [9], in a similar manner as small Cr clusters reported by Cheng *et al.* [10]. This unexpected finding is fortified by the fact that Zhang *et al.* used a plane wave basis set and Aguilera-Granja *et al*. used a localized orbital basis set and their results agreed with each other.

However, dimerization would be more difficult to imagine in Mo clusters, since Mo 4d valence electrons are located farther from the nucleus because 4d level must be orthogonal to 3d core level. Closely related to the question would be the ground state geometry of Mo trimer. Zhang *et al.* and Min *et al.* [11] reported an obtuse isosceles triangle as the ground state geometry of Mo trimer. However, in another DFT calculation, Pis Diez [12] predicted an equilateral triangle as the ground state geometry. Again, the results seem to be unrelated to the methodology.

Theoretical calculations of transition metal clusters via plane wave pseudopotential method are now widely used due to the rapid growth of computing capacity. We investigate the electronic properties and the structural properties of small Mo clusters using the projector augmented wave (PAW) pseudopotential method [13] and the generalized gradient approximation (GGA) [14-15]. Our results for the Mo dimer and bcc Mo crystal compare favorably with available experiments and theoretical calculations. When we used a pseudopotential that does not include the *4s-* and *4p-* semicore states explicitly, we obtained results that were close to the previous plane wave pseudopotential calculations by Zhang *et al.* [8]. However, we obtained quite different results when we used a pseudopotential that includes the *4s-* and *4p-* semicore states as the valence states. The most outstanding difference was that the dimerization tendency is greatly reduced.

## II. CALCULATION

The calculations were performed using the ABINIT package [16] with periodic boundary condition and gamma point sampling. We used projector augmented wave pseudopotential [13] with *4s-* and *4p-* semicore states [17]. Plane wave energy cutoff of 871 *eV* and double grid energy cutoff of 1633 *eV* were used. The box size was chosen as 15.9 Å ensuring that the total-energy converges within 1 *meV* . Exchange correlation energy was described by the Perdew-Burke-Ernzerhof parameterization within the generalized gradient approximation (PBE-GGA) [18]. Self-consistency cycles were repeated until the difference of the total energy becomes smaller than $2.7 \times 10^{-6} eV$ , twice in a row. The system was relaxed until the average force on the atoms becomes smaller than $3 \times 10^{-4} eV/\text{Å}$ by Broyden-Fletcher-Goldfarb-Shanno (BFGS) minimization scheme [19]. The spin occupations were determined by manually changing the target magnetic moment of the clusters and then minimizing the total energy [20].

## III. RESULTS AND DISCUSSION

Mo dimer has been studied as a testing ground of theoretical approaches [6-7]. In the early days, it was not clear enough if the density functional theory would be able to describe multiple bonding between transition metal atoms. The bond length from theoretical calculations varies from 1.65 Å to 2.1 Å, compared to 1.93 Å from experiments. We considered LSDA and GGA, both with and without the *4s-* and *4p-* semicore states. GGA does provide an improved description of Mo dimer over the LSDA. Also, the semicore correction achieves as much improvement within the LSDA and within the GGA. Mo dimer always had zero magnetic moment in our calculations. The results for Mo dimer are summarized in Table I. GGA with the semicore corrections achieves an excellent agreement with experiments. Without the semicore corrections, the bond lengths become smaller, the binding energy larger, and the vibration frequency higher.

We also considered the case of bcc Mo metal to understand how these two approaches would affect the more homogeneous system. The lattice constant, the cohesive energy, and the bulk modulus is calculated as 3.17 Å, 6.25 eV/atom, and 248 GPa, respectively, using the GGA with the semicore corrections, and as 3.17 Å, 6.30 eV/atom, and 269 GPa, respectively, without the semicore corrections. Experiments report 3.15 Å, 6.82 eV/atom, and 273 GPa [25]. Other experiments report a bulk modulus of 261 GPa [26]. Che *et al*. [27] reports a bulk modulus of 280 GPa from *ab initio* density functional theory calculations using norm-conserving pseudopotentials. The relaxation of the semicore provides an additional degree of freedom in the process of compression and thus decreases the calculated bulk modulus. Zhang *et al*. [8] mentions that a pseudopotential that explicitly includes *4p* electrons did not work well for bulk Mo. Since such anomaly is not observed in our case, the *4s* electrons must play an important role in the dynamics of Mo system. Dawaele *et al*. [17] also reported a significant improvement of the bulk elastic properties in GGA PAW calculations with semicore corrections. All these results suggest that the GGA with the s- and p- semicore corrections provides an excellent description of the energetics of Mo system, including small clusters.

There are also disagreements concerning the equilibrium structure of Mo trimer. The central question is whether there exists a strong tendency to form dimers as reported by Zhang *et al*. [8] and Aguilera-Granja *et al*. [9], in a similar manner of small Cr clusters reported by Cheng *et al*. [10] If such were true, we may envisage an acute isosceles triangular geometry as the ground state or as an excited state lying very close to the ground state. Also, in the artificial linear structures these authors studied, the dimerization tendency would be more vivid.

However, Pis Diez [12] and Lei [21] reported that the ground state of Mo trimer is an equilateral triangle. Aguilera-Granja *et al*. [9] predicted a linear structure as the ground state, but an acute isosceles triangle was lower in energy than an obtuse isosceles triangle. Zhang *et al*. [8] and Min *et al*. [11] reported an obtuse isosceles triangle as the ground state geometry. It is unusual that DFT calculations of transition metal cluster are at variance without apparent tendencies. In the present calculation, the ground state is always a triangular geometry. When the semicore corrections were not included, the ground state was an acute isosceles triangle with bond lengths 2.34 Å, 2.34 Å, 1.97 Å. However, with the semicore corrections, the equilibrium geometry becomes an obtuse isoceles triangle with bond lengths 2.17 Å, 2.17 Å, 2.42 Å. We note that the ground state structure of the Mo trimer depended very much on the plane wave cutoff energy, long after the total-energy of single Mo atom and the Mo crystal had converged within 1 meV. When we used 762 *eV* as the plane wave cutoff energy with the semicore corrections, the ground state geometry became an equilateral triangle of 2.22 Å. It seems possible that the variations in the published results are affected by the basis sets used in the calculations.

Also, the linear geometry showed a long bond to short bond ratio of 1.69 without semicore corrections, which again is close to 1.62 by Zhang *et al.* [8] and 1.80 by Aguilera-Granja *et al.* [9] But the ratio is reduced to 1.41 with semicore corrections. Thus, our results suggest a possibility that the strong dimerization tendency may have been exaggerated by an insufficient description of the core electron densities. Such insufficiency could originate from the absence of core electron densities and/or the incompleteness of the wave function basis sets, which in turn poses a possibility of an inadvertent error cancellation. The results for triangular Mo trimer are summarized in Table II, those for linear trimer in Table III. The triangular trimers have a magnetic moment of $2\,\mu_B$ per cluster, while the linear trimers have a magnetic moment of $6\,\mu_B$ per cluster. We note that our PAW results without semicore corrections are indeed very close to the ultrasoft pseudopotential plane wave calculation results by Zhang *et al.* [8]

The ground state of Mo tetramer is also very important in understanding the bonding properties of small clusters. For example, theoretical calculations predict a two-dimensional rhombic ground state for 3d transition metal tetramers such as $Cr_4$ [10] and $Cu_4$ [28], but a three-dimensional tetrahedral ground state for 4d transition metal tetramers such as $Pd_4$ [29]. Mo tetramer would also be a showcase of the strength of the dimerization tendency. Previous theoretical studies of Mo tetramer predict elongated tetrahedron structure as the ground state, basically two dimers separated by a larger distance [8, 11, 12]. The ratio of the long bond length with respect to the short bond length will reflect the strength of the dimerization tendency. Again, our results are similar to those of the previous plane wave calculations [8] when the semicore states are not included. The ground state is a three-dimensional elongated tetrahedron with two short dimer bonds of 1.84 Å and four long bonds of 2.91 Å with a magnetic moment of $2\mu_B$ per cluster and a binding energy of 2.97 eV/atom. Zhang *et al.* [8] predicts zero magnetic moment for the Mo tetramer. The two dimers are oriented perpendicular to each other, making the geometry fully three-dimensional. With the semicore states, the ground state geometry changes to a buckled square with four bonds of 2.22 Å, more or less resembling the two-dimensional rhombic ground state for 3d transition metal tetramers. This configuration has zero magnetic moment and a binding energy of 2.61 eV/atom.

The difference due to the treatment of the semicore states diminishes after n=5. Mo pentamer has a buckled square pyramid geometry with a binding energy of 3.08 eV/atom (without the semicore) and 2.92 eV/atom (with the semicore). Both has zero magnetic moment. The hexamer is a buckled square bipyramid with a binding energy of 3.36 eV/atom (without the semicore) and 3.25 eV/atom (with the semicore). Hexamer has no magnetic moment. The heptamer is a buckled pentagonal bipyramid. The binding energy without the semicore is 3.47 eV/atom and the binding energy with the semicore is 3.37 eV/atom. The magnetic moment is zero in both cases. The octamer geometry is a bicapped square bipyramid with a binding energy of 3.56 eV/atom (without the semicore) and 3.41 eV/atom (with semicore). The magnetic moment was $2\mu_B$ per cluster (without the semicore) and zero (with the semicore). The nonmagnetic configuration (without the semicore) exists very close to the magnetic configuration, within 15 meV/atom. The most stable geometries of Mo clusters are shown in Figure 1 (without the semicore) and Figure 2 (with the semicore).

The binding energies of the Mo clusters are shown as a function of the cluster size in Figure 3. Without the semicore states, the binding energy of Mo dimer is severely overestimated and the binding energy of Mo trimer quite underestimated, giving the illusion of a strong dimerization tendency in Mo cluster. It is more explicitly seen in the second-order difference of the binding energy (Figure 4), as calculated by $\Delta_2(n) = E(n+1) - 2E(n) + E(n-1)$. The even-odd oscillation attributed to the dimerization is drastically reduced.

The HOMO-LUMO gaps of the clusters are shown as a function of the cluster size in Figure 5. Although the difference of the total-energies calculated with and without semicore states gets smaller as the cluster size increases, there still exists a large difference in HOMO-LUMO gap at $n = 7$. As can be seen from the variation in the HOMO-LUMO gaps, the semicore states exert a far-reaching influence on the chemical activity of the clusters as well as the structural properties.

## IV. CONCLUSION

We studied the structural properties of small Mo clusters via plane wave projector augmented wave calculations. The *4s-* and *4p-* semicore states play an important role in the description of small Mo clusters and drastically reduce the dimerization tendency between Mo atoms. The semicore states had a far-reaching influence on the electronic and structural properties of small Mo clusters. Also, the determination of Mo cluster geometry seems to require much larger plane wave cutoff energy than the customary values taken in bulk or surface calculations.

## ACKNOWLEDGMENTS

The author is much indebted to the librarians of Daegu University and wishes to express special thanks for their great service. This research was supported in part by the Daegu University Research Funds.

Table I. Comparison of the bond length $d$ (Å), the binding energy $E$ ($eV$/cluster), and the vibration frequency $\omega$ ($cm^{-1}$) of Mo dimer.

|  | $d$ (Å) | $E$ ($eV$/cluster) | $\omega$ ($cm^{-1}$) |
|---|---|---|---|
| LSDA (without semicore) | 1.69 | -8.01 | 368 |
| LSDA(with semicore) | 1.74 | -5.52 | 353 |
| GGA(without semicore) | 1.93 | -5.81 | 425 |
| GGA(with semicore) | 1.94 | -3.79 | 432 |
| Delley *et al*. [6] | 1.95 | -4.35 | 520 |
| Bernholc *et al*. [7] | 2.1 | -4.2 | 360 |
| Zhang *et al*. [8] | 1.80 | -5.08 | - |
| Aguilera-Granja *et al*. [9] | 1.65 | -4.45 | - |
| Min *et al*. [11] | 1.97 | -4.79 | 542 |
| Pis Diez [12] | 1.97 | -5.34 | 552 |
| Lei [21] | 1.98 | -3.00 | 523 |
| experiment | 1.93 [22] | -4.2 [23] | 477 [22] |
| experiment | 1.94 [24] | -4.38 [24] | 477 [24] |

Table II. Comparison of the bond lengths $d$ (Å), the binding energy $E$ ($eV$/cluster), and the long bond length to the short bond length ratio of triangular Mo trimer. All calculations predict a magnetic moment of $2 \mu_B$ per cluster. Aguilera-Granja $et\ al.$ [9] predicts a linear structure as the ground state.

|  | $d$ (Å) | $E$ ($eV$/cluster) | Ratio |
|---|---|---|---|
| GGA(without semicore) | 1.97, 2.34, 2.34 | -7.14 | |
| GGA(with semicore) | 2.17, 2.17, 2.42 | -6.25 | 1.12 |
| Zhang $et\ al.$ [8] | 2.04, 2.04, 2.34 | -6.75 | 1.15 |
| Aguilera-Granja $et\ al.$ [9] | 1.93, 2.35, 2.35 | -5.13 | |
| Min $et\ al.$ [11] | 2.19, 2.19, 2.47 | -7.77 | 1.13 |
| Pis Diez [12] | 2.24 | -9.24 | 1.00 |
| Lei [21] | 2.29 | -5.13 | 1.00 |

Table III. Comparison of the bond lengths $d$ (Å), the binding energy $E$ ($eV$/cluster), and the long bond length to the short bond length ratio of linear Mo trimer. All calculations predict a magnetic moment of $6\,\mu_B$ per cluster. Aguilera-Granja *et al.* [9] predicts ground state.

|  | $d$ (Å) | $E$ ($eV$/cluster) | Ratio |
|---|---|---|---|
| GGA(without semicore) | 1.71, 2.90 | -6.47 | 1.69 |
| GGA(with semicore) | 1.99, 2.81 | -4.76 | 1.41 |
| Zhang *et al.* [8] | 1.80, 2.92 | -5.88 | 1.62 |
| Aguilera-Granja *et al.* [9] | 1.61, 2.89 | -5.89 | 1.80 |

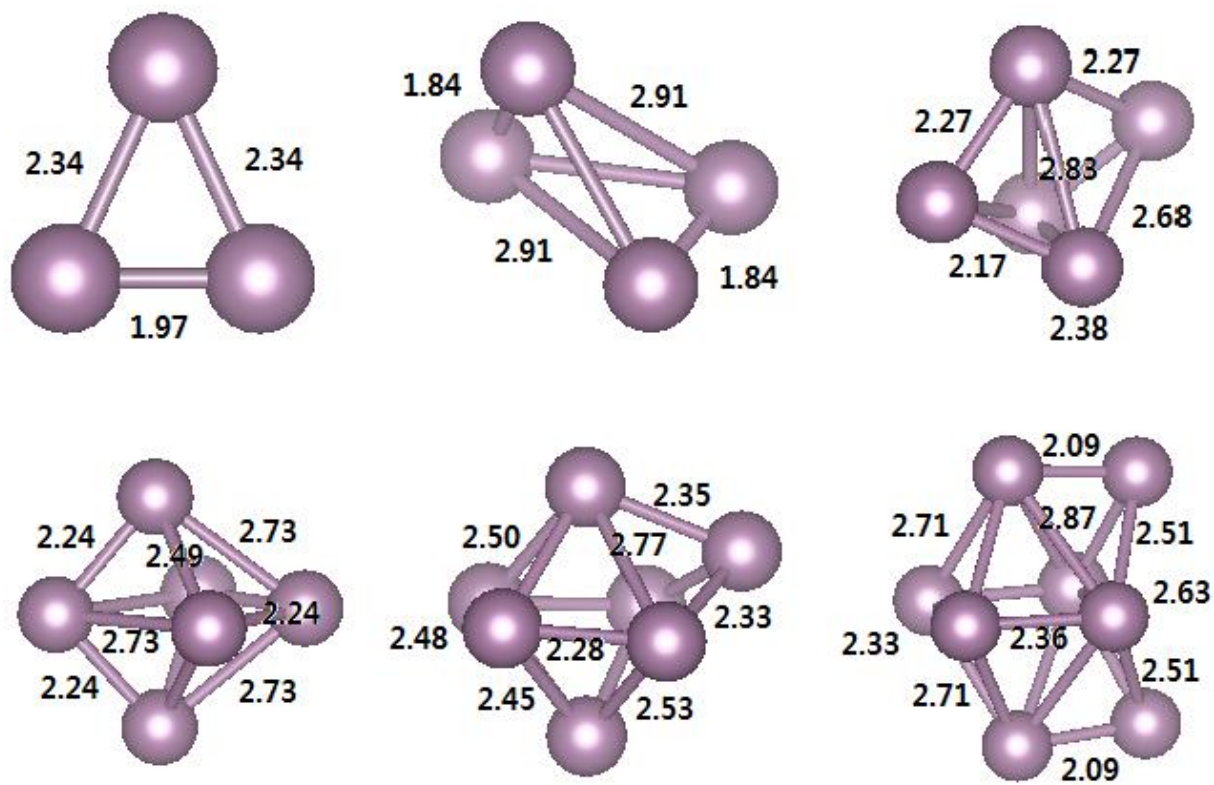

Fig.1

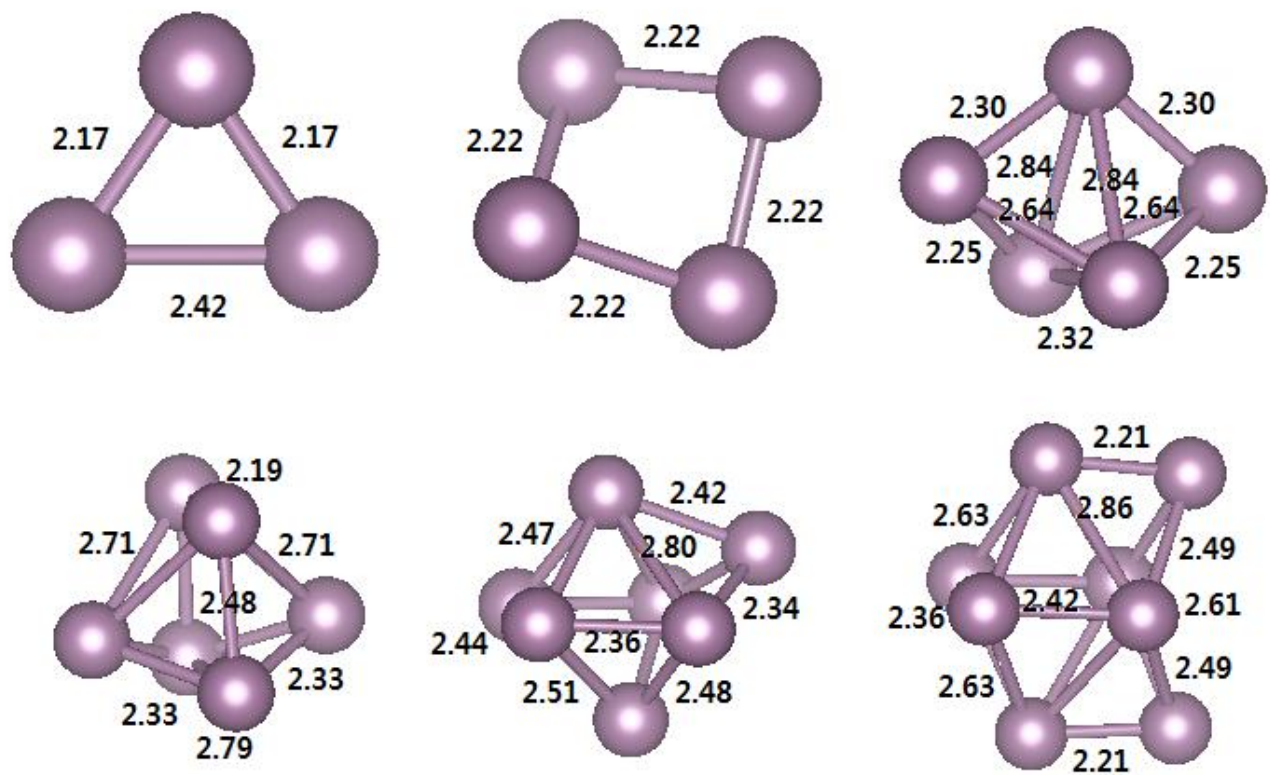

Fig.2

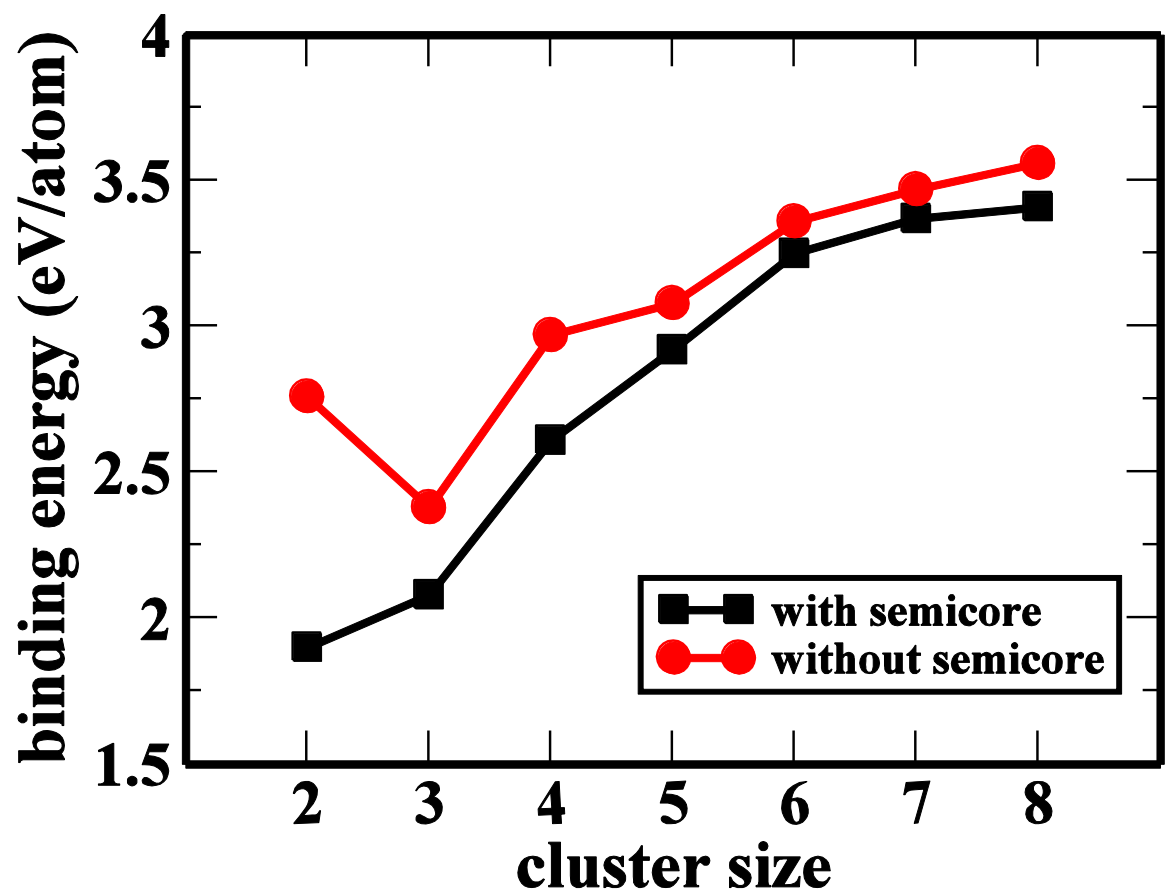

Fig. 3

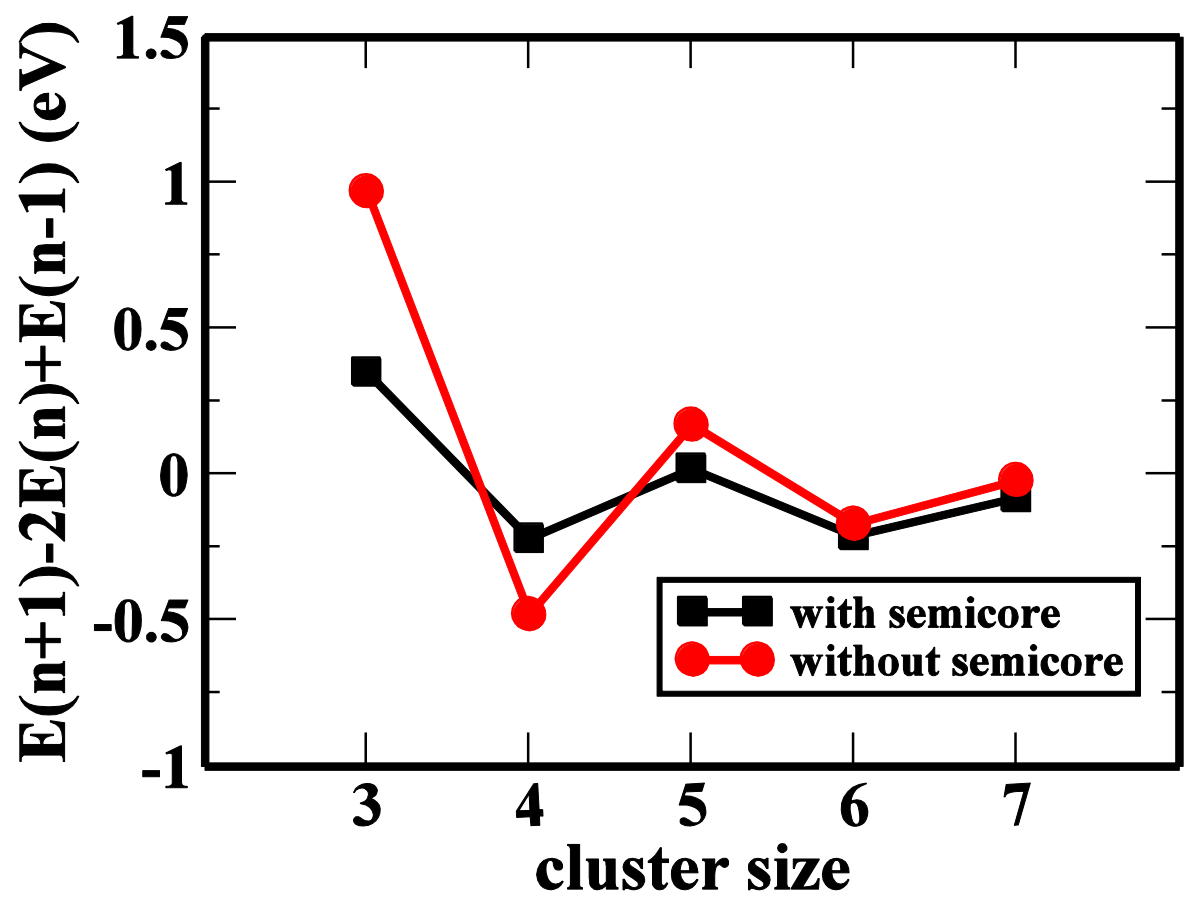

Fig. 4

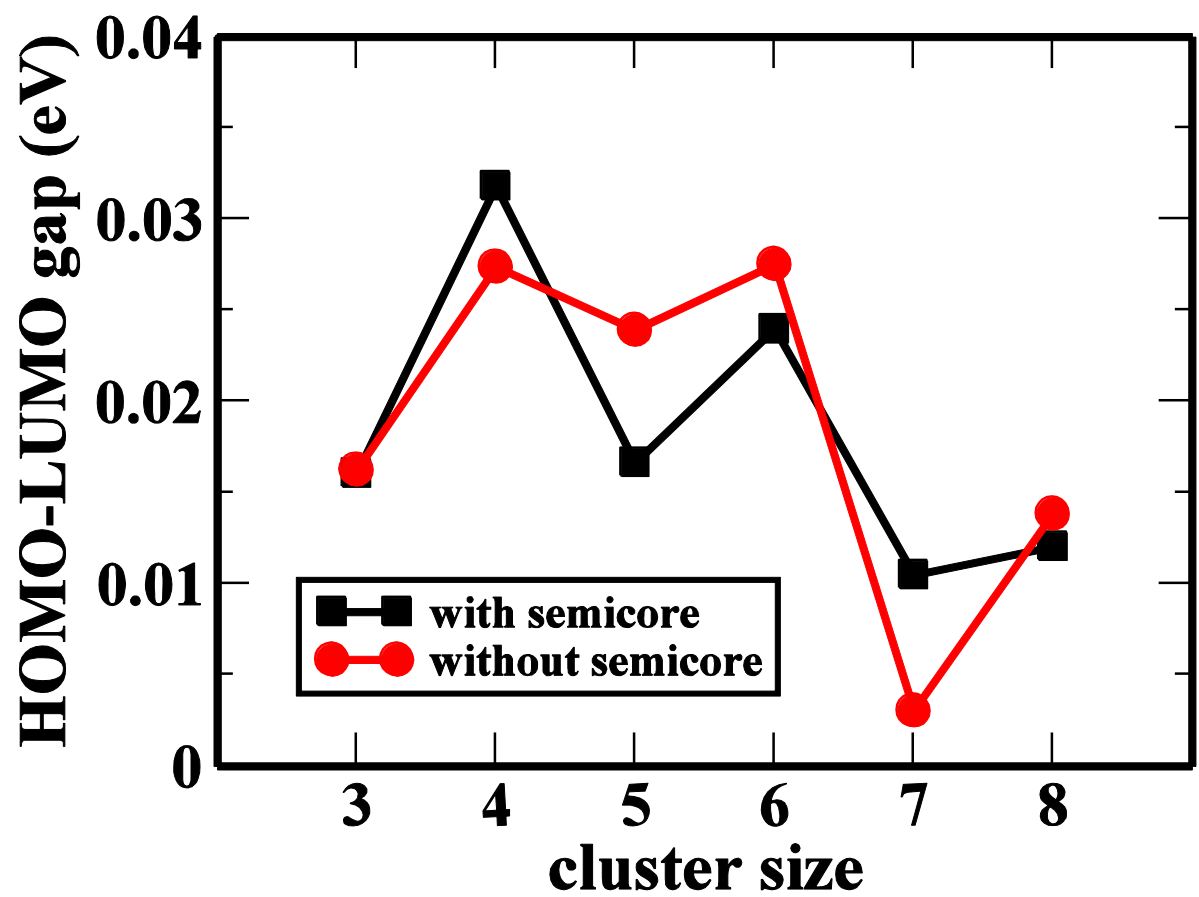

Fig. 5

Figure Captions.

Fig. 1. The most stable geometries of the Mo clusters obtained without the semicore states.

Fig. 2. The most stable geometries of the Mo clusters obtained with the semicore states

Fig. 3. The binding energy (eV/atom) of Mo cluster as a function of the cluster size. Without the semicore states, the binding energy of Mo dimer is severely overestimated and the binding energy of Mo trimer also quite underestimated, giving the illusion of a strong dimerization tendency in Mo cluster.

Fig. 4. The second-order difference of the binding energy as a function of the cluster size.

Fig. 5 The HOMO-LUMO gap as a function of the cluster size.